\pdfoutput=1
\documentclass[reprint,aps,pre,showpacs,twocolumn]{revtex4-1}
\usepackage{mathptmx} 
\usepackage{graphicx,color}
\usepackage{subfigure}
\usepackage{bm,amsmath,amssymb}

\newcommand{\eref}[1]{Eq.~(\ref{#1})}%
\newcommand{\Eref}[1]{Equation~(\ref{#1})}%
\newcommand{\fref}[1]{Fig.~\ref{#1}} %
\newcommand{\sref}[1]{Sec.~\ref{#1}}%
\newcommand{\Sref}[1]{Section~\ref{#1}}%
\newcommand{\sgn}[1]{\mathrm{sgn}({#1})}%
\newcommand{\erfc}{\mathrm{erfc}}%

\begin{document}

\title{Heat and work fluctuations for a harmonic oscillator}

\author{Sanjib Sabhapandit}

\affiliation{Raman Research Institute, Bangalore 560080, India}

\date{\today}



\begin{abstract}

The formalism of Kundu \textit{et al.} [J. Stat. Mech.  (2011)
P03007], for computing the large deviations of heat flow in harmonic
systems, is applied to the case of single Brownian particle in a
harmonic trap and coupled to two heat baths at different
temperatures. The large-$\tau$ form of the moment generating function
$\langle e^{-\lambda Q}\rangle\approx g(\lambda)\exp[\tau\mu
(\lambda)]$, of the total heat flow $Q$ from one of the baths to the
particle in a given time interval $\tau$, is studied and exact
explicit expressions are obtained for both $\mu(\lambda)$ and
$g(\lambda)$. For a special case of the single particle problem that
corresponds to the work done by an external stochastic force on a
harmonic oscillator coupled to a thermal bath, the large-$\tau$ form
of the moment generating function is analyzed to obtain the exact
large deviation function as well as the complete asymptotic forms of
the probability density function of the work.
\end{abstract}

\maketitle

\section{Introduction}

The emergence of the so-called \emph{fluctuation
relations}~\cite{Evans:93, Gallavotti:95, Jarzynski:97, Kurchan:98,
Lebowitz:99, Crooks:99, Hatano:01} has generated considerable
theoretical~\cite{Farago:02, vanZon:03, vanZon:04, Mazonka:99,
Narayan:04, Seifert:05, Baiesi:06, Bonetto:06, Visco:06, Harris:07,
Saito:07, Bodineau:04} and experimental~\cite{Wang:02, Wang:05,
Carberry:04, Goldburg:01, Feitosa:04, Garnier:05, Liphardt:02,
Collin:05, Majumdar:08, Douarche:06, Falcon:08, Bonaldi:09,
Ciliberto:10-2, Ciliberto:10} interests in studying fluctuation of
various stochastic quantities such as entropy production, heat flow,
particle transfer, power injection, work done, etc. in a given time
interval $\tau$, in nonequilibrium systems. In this context, one is
usually interested in the large-$\tau$ behavior of the probability
distribution of the studied stochastic quantity, say $Q$. The
probability distribution $P(Q)$ is expected to have a large deviation
form~\cite{Touchette:09} $P(Q)\sim \exp[\tau h(Q/\tau)]$ for large
$\tau$.  However, in spite of the large interests, there are only a
few systems where the large deviation function $h(q)$ has been
obtained exactly.

An interesting example is a free Brownian particle coupled to two heat
baths at different temperatures, initially introduced by Derrida and
Brunet~\cite{Derrida:05}.  In this model, the quantity of interest is
the heat flow from one of the baths to the particle in a given time
interval $\tau$.  It turns out that this model can be mapped to an
exactly solved problem, namely, the quantum harmonic oscillator ---
Visco~\cite{Visco:06} has obtained the exact characteristic function
$\langle e^{-\lambda Q}\rangle$ for all $\tau$, and thereafter, the
large deviation function $h(q)$ by analyzing the large-$\tau$ limit of
$\langle e^{-\lambda Q} \rangle$.

Now in addition, if a harmonic trap is introduced around the Brownian
particle, this seemingly simple model becomes quite non-trivial and
any methods for solving the relevant Fokker-Planck equation
[Eqs.~\eqref{FP-eq} and \eqref{FP-op}] is not known to us. Fortunately,
one does not require the complete solution of the Fokker-Planck
equation, as the large-$\tau$ behavior is essentially determined by
the largest eigenvalue and the corresponding eigenfunctions (left and
right) of the Fokker-Planck operator [see \eref{characteristic.1}]. It
is indeed remarkable that these functions could be found exactly (in
terms of some integrals) for a harmonic chain whose two ends are
coupled to heat baths at different temperatures~\cite{Kundu:11}. The
harmonically bound Brownian particle is a special case of the harmonic
chain. In this paper we find the eigenvalue and the eigenfunctions
explicitly for the harmonically bound particle, and using those we
then find the characteristic function $\langle e^{-\lambda Q}\rangle$
for large $\tau$.
In fact, a special case of this model --- that concerns the
fluctuations of the work done by an external stochastic force on a
harmonic oscillator coupled to a thermal bath --- has been realized in
a recent experiment~\cite{Ciliberto:10}.  
The goal of the this paper is to analyze this problem in detail, as
the methods should be useful for other similar problems. Some of the
main results have been reported in~\cite{SS:11}.

This paper is organized as follows. In \sref{Brownian particle} we
discuss the problem of the harmonically bound Brownian particle
coupled to two heat baths. \Sref{driven HO} contains the special case
of the harmonic oscillator driven by a random force.  In \sref{LDF} we
obtain the large deviation function of the work fluctuation and
in \sref{finite time} we find the complete asymptotic form of the
probability density function of the work.  \Sref{stochastic
integration} contains some remarks on the evaluation of stochastic
integrals that define heat, work, etc., in numerical simulations or
from experimental data.  Finally we summarize in \sref{summary}. Some
of the details are presented in the appendices. In
Appendix~\ref{Harmonic chain} we outline the relevant results of
Ref.~\cite{Kundu:11} for the harmonic chain and Appendix~\ref{N=1
case} contains some of the details of the bounded Brownian particle
case discussed in \sref{Brownian particle}.  In Appendix~\ref{uniform
asymptotic} we outline the method of uniform asymptotic expansion of a
integral having a saddle-point near a pole,  which is used to obtain
the complete asymptotic form of the probability density function
in \sref{finite time}.

\section{Brownian particle in a harmonic potential coupled to  two thermostats}
\label{Brownian particle}

Consider a Brownian particle of mass $m$ in a one-dimensional harmonic
potential with spring constant $k$ and coupled to two white noise
Langevin thermal baths at two different temperatures $T_L$ and $T_R$
respectively. The displacement $x(t)$ from its mean position and the
velocity $v(t)$ of the particle are described by the equations
\begin{equation}
\dot{x}=v \quad\text{and}\quad m\dot{v} = -\gamma v -k x + \eta(t),
\label{eqm-s-bpart}
\end{equation}
where $\gamma=\gamma_L +\gamma_R$ and $\eta(t)=\eta_L(t) + \eta_R(t)$
with $\eta_L$ and $\eta_R$ being the Gaussian white noises with mean
zero and correlators:
\begin{align}
\label{correlator-1}
&\bigl\langle \eta_L(s) \eta_L(t)\bigr\rangle = 2
d_L \delta(s-t) \quad\text{with}~d_L=\gamma_L k_B T_L,\\
\label{correlator-2}
&\bigl\langle \eta_R(s) \eta_R(t)\bigr\rangle = 2 d_R 
\delta(s-t) \quad\text{with}~d_R=\gamma_R k_B T_R,\\
\label{correlator-3}
&\text{and}\quad\bigl\langle \eta_L(s) \eta_R(t)\bigr\rangle = 0,
\end{align}
where $k_B$ being the Boltzmann constant. The quantity of interest is
the total amount of heat flowing from the reservoir at temperature
$T_L$ to the particle in a time duration $\tau$:
\begin{equation}
  Q=\int_0^\tau \bigl[\eta_L(t) - \gamma_L v(t)\bigr] v(t)\; d t.
\label{heat}
\end{equation}
This system is a special case ($N=1$) of the harmonic chain (of $N$
particles) connected at its two ends to reservoirs at different
temperatures, that has been studied recently~\cite{Kundu:11} and is
outlined in Appendix~\ref{Harmonic chain}.

We consider the restricted characteristic function
\begin{equation}
Z(\lambda,x,v,\tau|x_0,v_0)=\bigl \langle e^{-\lambda
Q}\, \delta[x-x(\tau)] \delta[v-v(\tau)] \bigr\rangle_{(x_0,v_0)}
\label{restricted CF}
\end{equation}
for fixed initial and final configurations, $(x_0,v_0)$ and $(x,v)$
respectively. It satisfies the Fokker-Planck equation~\cite{Kundu:11}
\begin{equation}
\frac{\partial}{\partial \tau} Z(\lambda,x,v,\tau|x_0,v_0)
= \mathcal{L}_\lambda Z(\lambda,x,v,\tau|x_0,v_0)
\label{FP-eq}
\end{equation}
with the initial condition
$Z(\lambda,x,v,0|x_0,v_0)= \delta(x-x_0)\, \delta(v-v_0)$ and the
Fokker-Planck operator is given by
\begin{multline}
\mathcal{L}_\lambda=\frac{d_L+d_R}{m^2}\frac{\partial^2}{\partial v^2}
+
\biggl[\frac{k}{m} x + \frac{\gamma+ 2\lambda d_L}{m} v \biggr]
\frac{\partial}{\partial v} \\
-v\frac{\partial}{\partial x}
+\lambda(\gamma_L + \lambda d_L) v^2
+\frac{\gamma + \lambda d_L}{m}.
\label{FP-op}
\end{multline}

The solution of the Fokker-Planck equation can be formally expressed
in the eigenbases of the operator $\mathcal{L}_\lambda$ and the large
$\tau$ behavior is dominated by the term having the largest
eigenvalue. Thus, for large $\tau$,
\begin{equation}
  Z(\lambda,x,v,\tau |x_0,v_0) 
\sim
  \chi(x_0,v_0,\lambda)\Psi(x,v,\lambda)\,
  e^{\tau\mu(\lambda)},
  \label{characteristic.1}
\end{equation}
where $\Psi(x,v,\lambda)$ is the eigenfunction corresponding to the
largest eigenvalue $\mu(\lambda)$ and $\chi(x_0,v_0,\lambda)$ is the
projection of the initial state onto the eigenstate corresponding to
the eigenvalue $\mu(\lambda)$. These functions are obtained in
Appendix~\ref{N=1 case} and we find that
\begin{align}
\label{mu(lambda) N=1}
&\qquad\mu(\lambda)=\frac{1}{2\tau_\gamma}\bigl[1-\eta(\lambda) \bigr], \\
\label{psi}
&\Psi(x,v,\lambda)=\left[\frac{\gamma\eta(\lambda)\sqrt{km}}
{2\pi(d_L+d_R)}\right]
\exp\bigl[-B_+(\lambda) E(x,v)\bigr],\\
\label{chi}
&\text{and}\quad\chi(x_0,v_0,\lambda)= 
\exp\bigl[-B_-(\lambda) E(x_0,v_0)\bigr],
\end{align}
where $\tau_\gamma = m/\gamma$ is the viscous relaxation time,
\begin{align}
\label{eta(lambda)}
&\eta(\lambda)=\sqrt{1+4\frac{d_L
d_R}{\gamma^2}\,\lambda(\Delta\beta-\lambda)},\\
&E(x,v)=\frac{1}{2} k x^2 + \frac{1}{2} m v^2, 
\end{align}
and $B_\pm(\lambda)$ is given by \eref{B_pm}. Fogedby and Imparato
have recently shown~\cite{Fogedby:11} that $\mu(\lambda)$ can also be
obtained by the Derrida-Brunet method~\cite{Derrida:05}.

Using the explicit forms one can easily verify that
$\mathcal{L}_\lambda \Psi (x,v,\lambda)
= \mu(\lambda) \Psi(x,v,\lambda)$. Moreover, since from \eqref{B_pm}
we get $B_+(\lambda) + B_-(\lambda)= \gamma \eta(\lambda)/ (d_L+d_R)$,
it immediately follows that
\begin{equation}
\int_{-\infty}^\infty\int_{-\infty}^\infty
\chi(x,v,\lambda) \Psi(x,v,\lambda)\, dx\, dv =1,
\end{equation}
which is demanded by the normalization.  From the above expressions,
we also find that $\mu(0)=0$ and $\chi(x_0,v_0,0)=1$.  Since
$\lambda=0$ case of \eref{restricted CF} gives the probability
distribution of the phase-space variables and $\mu(0)$ is the largest
eigenvalue, it follows from \eref{characteristic.1} that $\Psi(x,v,0)$
is the steady-state distribution of the phase-space.
Therefore, averaging over the initial variables $(x_0,v_0)$ with
respect to $\Psi(x_0,v_0,0)$ and integrating over the final variables
$(x,v)$, we find the characteristic function of the heat flow in the
steady state as
\begin{equation}
Z(\lambda,\tau) =\bigl \langle e^{-\lambda Q}\bigr\rangle
\approx g(\lambda)\,   e^{\tau\mu(\lambda)},
\label{Z-asymptotic}
\end{equation}
where
\begin{equation}
g(\lambda)=
\frac{4\eta(\lambda)}
{\bigl[1+\eta(\lambda)\bigr]^2 - \bigl[2 \lambda d_L/\gamma\bigr]^2}.
\label{g(lambda) N=1}
\end{equation}

Interestingly, both $\mu(\lambda)$ and $g(\lambda)$ are independent of
the spring constant $k$. However, while $\mu(\lambda)$ is same for
both $k\not=0$ and $k=0$ cases (the latter was obtained in
Ref.~\cite{Visco:06}), $g(\lambda)$ for $k\not=0$ differs from that
for $k=0$. More precisely, $g(\lambda)|_{k\not=0} =
[g(\lambda)|_{k=0}]^2$. The $k\rightarrow 0$ limit of $g(\lambda)$
is not same as the $k=0$ case. Therefore, although the large deviation
function are same for both the cases, the precise asymptotic form of
the probability density functions of $Q$ are different.

The leading behavior of the probability density function
$P(Q)\sim \exp[\tau h(Q/\tau)]$ can be obtained by
inverting \eref{Z-asymptotic} using the saddle-point approximation.
Usually the prefactor $g(\lambda)$ can be ignored in such calculation
and the large deviation function $h(q)$ is related to
$\mu(\lambda)$ by the Legendre transform
\begin{equation}
h(q)= \mu(\lambda^*) +\lambda^* q, 
\qquad
-\mu'(\lambda^*)=q.
\end{equation}
However, if $g(\lambda)$ has any singularities in the region of the
saddle-point integration, the functions $h(q)$ and $\mu(\lambda)$ are
not simply related by the Legendre transform, and it it important to
retain the prefactor $g(\lambda)$ in the saddle-point
calculation~\cite{Farago:02, vanZon:03, Visco:06}, as we see in the
next section where we consider a special case that corresponds to an
experiment reported in Ref.~\cite{Ciliberto:10}.

\section{Harmonic oscillators  driven by an external random force}
\label{driven HO}

Consider a harmonic oscillator coupled to a thermal bath and driven
out of equilibrium by an external Gaussian random force.  The
displacement $x(t)$ of the harmonic oscillator from its mean position
is described by the Langevin equation
\begin{equation}
m\frac{d^2x}{dt^2}+\gamma \frac{dx}{dt} +k x =\zeta_T (t) +f_0 (t),
\label{Langevin-HO}
\end{equation}
where $m$ is the mass, $\gamma$ is the viscous drag coefficient and
$k$ is the spring constant.  The interaction with the thermal bath is
modeled by a Gaussian white noise $\zeta_T(t)$ with zero-mean
$\langle \zeta_T(t)\rangle=0$.  The externally applied force $f_0(t)$
is again a Gaussian random variable with $\langle f_0(t) \rangle =0$,
and $\zeta_T$ and $f_0$ are uncorrelated.  \Eref{Langevin-HO} is
asymmetric in $\zeta_T$ and $f_0$ --- the fluctuation-dissipation
theorem relates the thermal fluctuation to the viscous drag as
$\langle \zeta_T(s) \zeta_T(t) \rangle= 2 D \delta(s-t)$ where
$D=\gamma k_B T$ with $T$ being the temperature of the bath and $k_B$
being the Boltzmann constant, whereas the fluctuation of the external
force $\langle f_0(s) f_0(t) \rangle = (\delta f_0)^2 \delta (s-t)$ is
independent of $\gamma$. The quantity of interest is the work done by
the external random force $f_0(t)$ on the harmonic oscillator in a
time interval $\tau$, in the nonequilibrium steady state. This is
given (in units of $k_B T$) by
\begin{equation}
W_\tau=\frac{1}{k_B T}\int_0^\tau  f_0(t) \frac{dx}{dt} \, dt,
\label{Work}
\end{equation}
with the initial condition (at $\tau = 0$) drawn from the steady state
distribution.

It is evident that this harmonic oscillator problem can be mapped to
the problem of the Brownian particle discussed in the previous
section, with the following set of transformations:
\begin{itemize}
\item 
$\eta_L(t) \rightarrow f_0(t)$ with $\gamma_L \rightarrow 0,\,
T_L\rightarrow \infty$ while keeping $\gamma_L T_L=d_L$ fixed and
$d_L\rightarrow (\delta f_0)^2/2$.

\item
$\eta_R(t) \rightarrow \zeta_T(t)$ with $T_R\rightarrow T$,
$\gamma_R\rightarrow \gamma$, and $d_R\rightarrow D$.
\end{itemize}
Under these transformations, we have
$Q\rightarrow (\gamma/D)^{-1} W_\tau$.  Therefore,
using \eref{Z-asymptotic} we now get
\begin{equation}
\bigl \langle e^{-\lambda W_\tau}\bigr\rangle
\approx g(\lambda)\,   e^{\tau\mu(\lambda)},
\end{equation}
where $\mu(\lambda)$ and $g(\lambda)$ are given by \eref{mu(lambda)
N=1} and \eref{g(lambda) N=1} respectively with the transformation
$\lambda \rightarrow (\gamma/D)\, \lambda$.

It is useful to express the relative strength of the external force
with respect to the thermal noise in terms of the dimensionless
parameter
\begin{equation}
\alpha=
\frac{(\delta f_0)^2}{2D}=
\frac{\langle x^2\rangle}{\langle x^2\rangle_\text{eq}}-1, \quad\text{and}~
\alpha \in (0,\infty),
\end{equation}
where $\langle x^2\rangle$ and $\langle x^2\rangle_\text{eq}$ are the
variance of $x$ in the nonequilibrium steady state (for $f_0 \not= 0$)
and in equilibrium (for $f_0=0$) respectively.  Using this parameter
$\alpha$ and the above transformations, the expression of
$\eta(\lambda)$ given by \eqref{eta(lambda)} becomes
\begin{equation}
\eta(\lambda)=\sqrt{1+4\alpha\lambda(1-\lambda)}.
\end{equation}
We also rewrite $g(\lambda)$ as
\begin{equation}
g(\lambda)=
\frac{2}
{1+\eta(\lambda) -2\alpha\lambda}
\times
\frac{2\eta(\lambda)}
{1+\eta(\lambda) +2\alpha\lambda},
\label{g(lambda)}
\end{equation}
where, the first factor in the above equation is due to the averaging
over the initial conditions with respect to the the steady state
distribution and the second factor is due to the integrating out of
the final degrees of freedom.

The probability density function of the work done is related to its characteristic function by
the inverse Fourier transform
\begin{equation}
P(W_\tau=w\tau) \approx \frac{1}{2\pi i} \int_{-i\infty}^{+i\infty} g(\lambda)\,
e^{\tau f_w(\lambda)}\, d\lambda,
\label{PDF}
\end{equation}
where 
\begin{equation}
f_w(\lambda)= \frac{1}{2}\bigl[1-\eta(\lambda) \bigr]
+\lambda w
\label{f_w}
\end{equation}
and we have set $\tau_\gamma=1$ --- this is equivalent to redefining the
time in the unit of $\tau_\gamma$, i.e., $\tau/\tau_\gamma \rightarrow
\tau$.  The integration is done along the
imaginary axis (vertical contour through the origin) in the
complex-$\lambda$ plane.

\subsection{The large deviation function}
\label{LDF}

\begin{figure}
\includegraphics[width=\hsize]{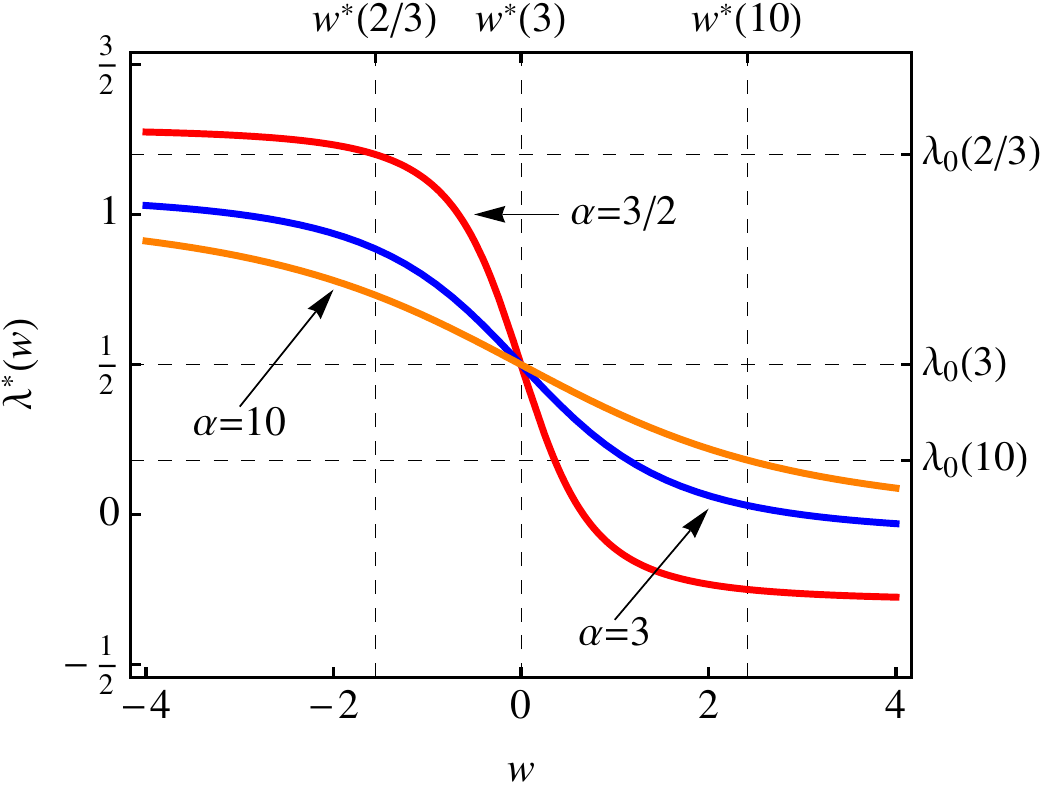} 
\caption{\label{lambda*-figure}(Color
  online).  Plots of the saddle-point $\lambda^*$ as a function of $w$
 for $\alpha=3/2$ (red), $3$ (blue) and $10$ (orange). The horizontal
 dashed lines mark the positions of the pole $\lambda_0$. The vertical
 dashed lines mark $w^*$ for respective values of $\alpha$.}
\end{figure}

\begin{figure}
\includegraphics[width=.95\hsize]{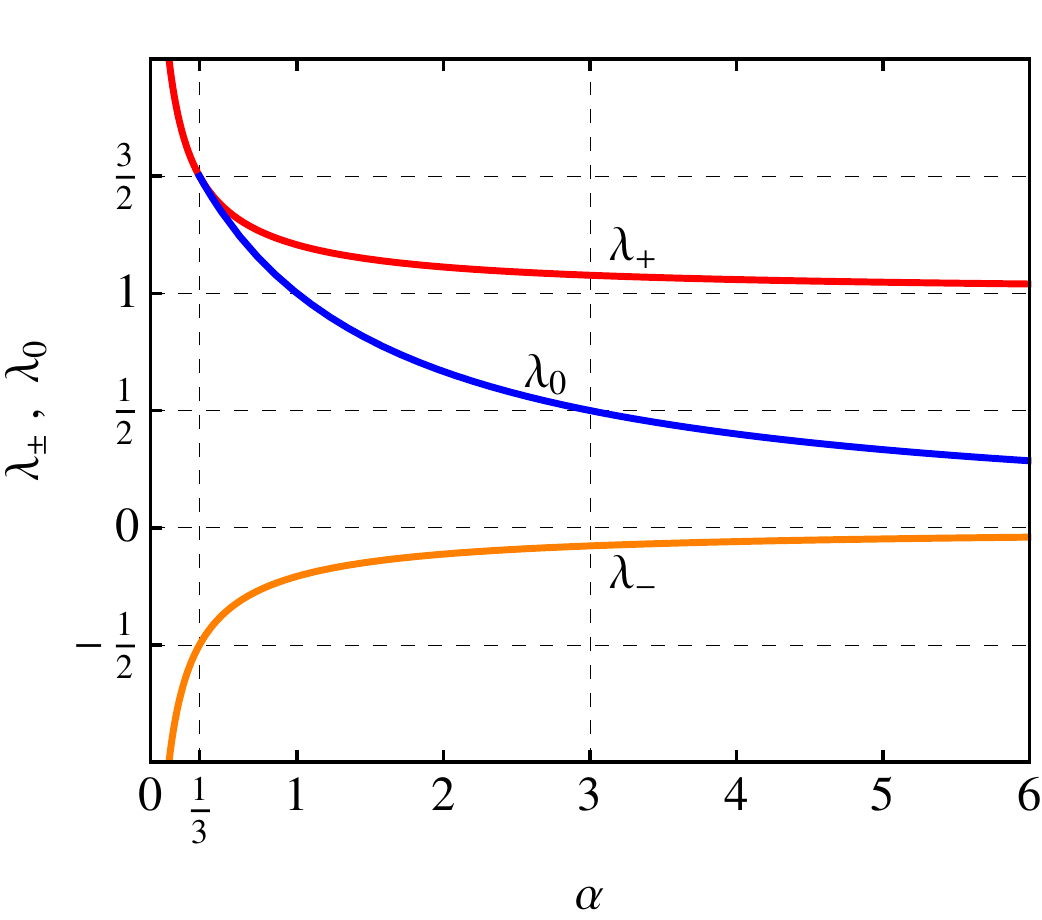} 
\caption{\label{lambda-figure}(Color
  online). $\lambda_+$, $\lambda_-$, and $\lambda_0$ are plotted (in
 red, orange, and blue colors, respectively) against
 $\alpha$.}  
\end{figure}

The large-$\tau$ behavior of $P(W_\tau)$ can be obtained from the
saddle point approximation of the integral in \eref{PDF}.  The
saddle-point $\lambda^*$ is obtained from the solution of the
condition $f'_w(\lambda^*)=0$ as
\begin{equation}
\lambda^*(w) =\frac{1}{2}
\left[1-\frac{w}{\sqrt{w^2+\alpha}}\sqrt{1+\frac{1}{\alpha}} \right].
\label{lambda*}
\end{equation}
It follows from the above expression that $\lambda^*(w)$ is a
monotonically decreasing function of $w$ (see \fref{lambda*-figure})
and $\lambda^*(w\rightarrow\mp \infty)\rightarrow\lambda_\pm$, where
\begin{equation}
\lambda_\pm =\frac{1}{2}\left[1\pm \sqrt{1+\frac{1}{\alpha}} \right].
\label{lambda_pm}
\end{equation}
Therefore, $\lambda^*\in (\lambda_-,\lambda_+)$. The $\alpha$
dependence of $\lambda_\pm$ are displayed in \fref{lambda-figure}.  We
note that $\eta(\lambda)$ can be written in terms of $\lambda_\pm$ as
\begin{equation}
\eta(\lambda)=\sqrt{4\alpha(\lambda_+-\lambda)(\lambda-\lambda_-)}.
\end{equation}
Clearly, $\eta(\lambda)$ has two branch points on the real-$\lambda$
line at $\lambda_\pm$, and $\eta(\lambda)$ is real and positive for
$\lambda\in(\lambda_-,\lambda_+)$. Therefore, $f_w(\lambda)$ is also
real on the real-$\lambda$ line in the interval
$(\lambda_-,\lambda_+)$. At $\lambda=\lambda^*$ we find
\begin{equation}
\eta(\lambda^*)=\frac{\sqrt{\alpha(1+\alpha)}}{\sqrt{w^2+\alpha}}.
\label{eta-lambda-star}
\end{equation}

Let us now look at the analyticity of $g(\lambda)$ given
by \eref{g(lambda)} for $\lambda\in (\lambda_-,\lambda_+)$.  It is
clear that $g(\lambda)$ is real for $\lambda\in
(\lambda_-,\lambda_+)$.  From \eref{lambda_pm} we find that
\begin{equation}
2\alpha\lambda_- + 1  = \sqrt{1+\alpha} \bigl[\sqrt{1+\alpha}
-\sqrt{\alpha}\bigr] >0.
\end{equation}
Therefore, $1+\eta(\lambda) + 2 \alpha \lambda >0$ for $\lambda\in
(\lambda_-,\lambda_+)$ for all $\alpha \in (0,\infty)$.
Again, from \eref{lambda_pm} we find that
\begin{equation}
2\alpha\lambda_+ -1 = \sqrt{\alpha (1+\alpha)} -(1-\alpha).
\end{equation}
The right side of the above equation is negative for $\alpha <
1/3$. Thus, $1-2\alpha \lambda >0$ for $\lambda \le \lambda_+$ and
$\alpha < 1/3$.  Consequently, when $\alpha <1/3$, we have
$1+\eta(\lambda) - 2 \alpha \lambda >0$ for $\lambda\in
(\lambda_-,\lambda_+)$.  Therefore, $g(\lambda)$ does not have any
singularities in the interval $(\lambda_-,\lambda_+)$ as long as
$\alpha < 1/3$.  Ignoring the subleading contribution $g(\lambda)$ in
the saddle-point calculation gives
\begin{equation}
P(W_\tau=w\tau)\sim e^{\tau h_1(w)}, 
\end{equation}
with
\begin{equation}
h_1(w)= f_w(\lambda^*).
\label{h1(w)-0}
\end{equation}
By substituting the expression of $\lambda^*$ in $f_w(\lambda^*)$,
after some algebra, we find that
\begin{equation}
h_1(w)=\frac{1}{2}\left[1+w
- \sqrt{w^2+\alpha}\sqrt{1+\frac{1}{\alpha} }\right].
\label{h1(w)}
\end{equation}

We now consider the case $\alpha \ge 1/3$.  In this case, it is useful
to express $g(\lambda)$ in the form
\begin{equation}
g(\lambda)=-
\frac{\eta(\lambda)\, [\eta(\lambda) + 2\alpha\lambda -1]}
{\alpha (1+\alpha)\, \lambda\,[1+\eta(\lambda) +2\alpha\lambda]}
\,\frac{1}{\lambda-\lambda_0}~,
\label{g(lambda)-2}
\end{equation}
where
\begin{equation}
\lambda_0=\frac{2}{1+\alpha}.
\label{lambda0}
\end{equation}
and $\lambda_- < 0 <\lambda_0 \le \lambda_+ $
(see \fref{lambda-figure}).  In order for $\lambda_0$ to be a zero of
$1+\eta(\lambda) - 2 \alpha \lambda$, it has to satisfy the condition
$2\alpha\lambda_0 -1 = \eta(\lambda_0) \ge0$. Using the above
expression of $\lambda_0$ we get
\begin{equation}
\eta(\lambda_0) = 2\alpha\lambda_0-1 = \frac{3\alpha -1 }{1+\alpha}.
\end{equation}
Therefore, $g(\lambda)$ possesses a simple pole at $\lambda_0$ when
$\alpha \ge 1/3$.  As shown in \fref{lambda*-figure}, for any given
$\alpha$, as $w$ decreases from $+\infty$ to $-\infty$, the saddle
point $\lambda^*(w)$ moves unidirectionally from $\lambda_-$ to
$\lambda_+$ on the real $\lambda$ line.  For sufficiently large $w$,
we have $\lambda_- < \lambda^* <\lambda_0$. In such a situation, the
contour of integration can be deformed smoothly through the saddle
point $\lambda^*$ and therefore we still have $P(W_\tau=w\tau)\sim
e^{\tau h_1(w)}$.  However, as one decreases $w$, at some particular
value $w=w^*$, the saddle-point hits the singularity --- where $w^*$
is found by solving $\lambda^*(w^*)=\lambda_0$ as
\begin{equation}
w^*=\frac{\alpha(\alpha-3)}{3\alpha-1}.
\label{w*}
\end{equation}
For $w<w^*$, we then have $0 < \lambda_0 <\lambda^*$. Therefore, while
shifting the contour of integration in the complex-$\lambda$ plane,
from its original path along the imaginary-$\lambda$ axis to the
steepest descent path through the saddle-point $\lambda^*$, a
contribution from the pole is picked up according to the residue
theorem, which to the leading order is $\exp[{\tau f_w(\lambda_0)}]$.
Now the second derivative of $f_w(\lambda)$ along the real-$\lambda$
axis at $\lambda^*$ can be found to be
\begin{equation}
f''_w (\lambda^*)= \frac{2
(w^2+\alpha)^{3/2}}{\sqrt{\alpha(1+\alpha)}}.
\label{f2}
\end{equation}
The above expression is always positive, which means that
$f_w(\lambda)$ has a minimum at $\lambda^*$ along real-$\lambda$.
That implies $f_w(\lambda^*) < f_w(\lambda_0)$.  Consequently, the
leading saddle-point contribution $\exp[{\tau f_w(\lambda^*)}]$ is
smaller than the leading pole contribution.  Therefore, the leading
behavior of $P(W_\tau)$ is given by
\begin{equation}
P(W_\tau=w\tau)\sim e^{\tau h_2(w)}, 
\end{equation}
with
\begin{equation}
h_2(w)= f_w(\lambda_0).
\label{h2(w)-0}
\end{equation}
By substituting $\lambda_0$ in $f_w(\lambda_0)$, after some algebra, we
find that
\begin{equation}
h_2(w)=\frac{1-\alpha}{1+\alpha} + \frac{2w}{1+\alpha}.
\label{h2(w)}
\end{equation}

To summarize, the large deviation function, defined by
\begin{equation}
h(w)=\lim_{\tau\rightarrow\infty}\frac{1}{\tau} \ln P(W_\tau=w\tau),
\end{equation}
is given by the following.  For $\alpha < 1/3$:
\begin{equation}
h(w)=h_1(w),
\label{form1}
\end{equation}
and 
for $\alpha \ge 1/3$:
\begin{equation}
h(w)=
\begin{cases}
h_1(w)
 & \text{for}~ w \ge w^*,\\
h_2(w) & \text{for}~ w \le w^*,
\end{cases}
\label{form2}
\end{equation}
where $h_1(w)$ and $h_2(w)$ are given by \eref{h1(w)}
and \eref{h2(w)}, respectively, and $w^*$ is given by \eref{w*}. It is
easy to show that $h_1(w^*)=h_2(w^*)$ and $h'_1(w^*)=h'_2(w^*)$.

\subsection{Finite time corrections}
\label{finite time}

\begin{figure}
\includegraphics[width=.95\hsize]{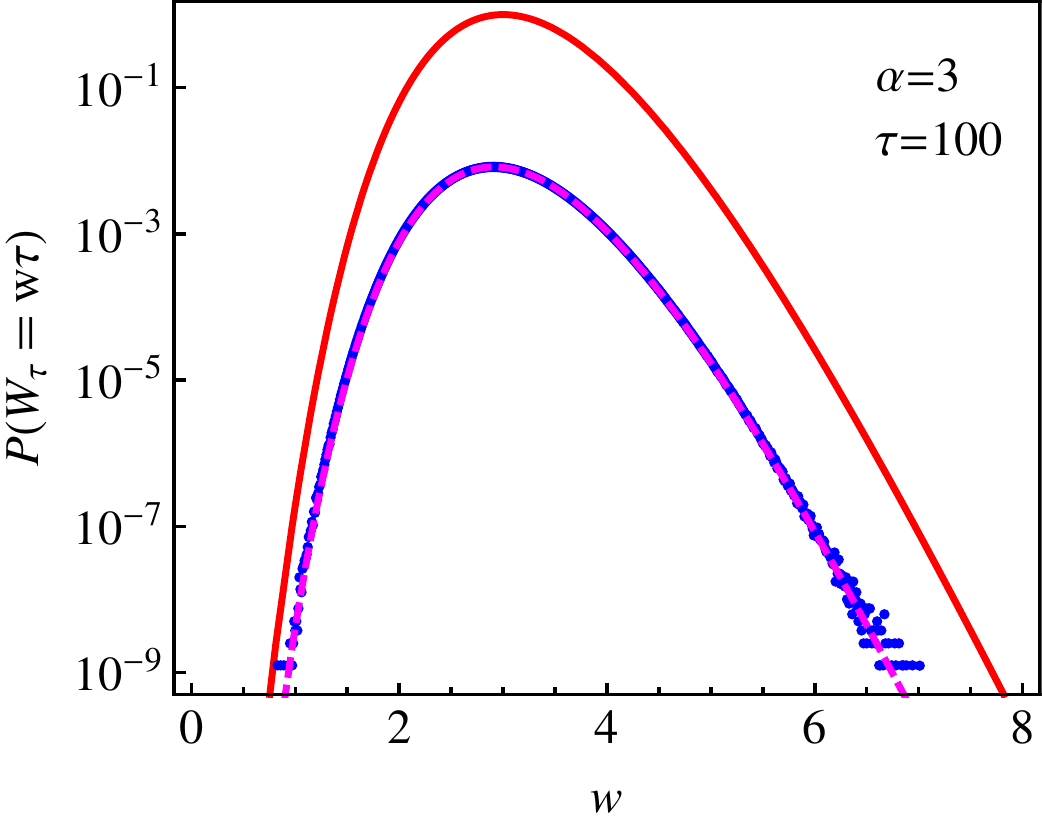} 
\caption{\label{ldcomp-figure}(Color
  online). $P(W_\tau)$ against the scaled variable $w=W_\tau/\tau$ for
  $\tau=100$ and $\alpha=3$. The points (blue) are obtained from
  numerical simulation, and the solid line (red) plots the large
  deviation form given by \eref{LDform}. The dashed (magenta) line
  plots the asymptotic form given by \eref{PDF-1}.}
\end{figure}

In the previous subsection we have obtained the leading behavior of
$P(W_\tau)$, which has the large deviation form
\begin{equation}
P(W_\tau=w\tau)\sim e^{\tau h(w)},
\label{LDform}
\end{equation}
with the large deviation function given by either \eref{form1}
or \eref{form2} depending on whether $\alpha < 1/3$ or $\alpha \ge
1/3$.  In \fref{ldcomp-figure} we plot this form together with
$P(W_\tau)$ that we have obtained from numerical simulation for
$\tau=100$.  It is evident from the comparison that the large
deviation form given by \eref{LDform} is not adequate to explain
simulation data (or experimental data) of finite time. In this
subsection we obtain the sub-leading contributions to the large
deviation form.

\begin{figure}
\includegraphics[width=.9\hsize]{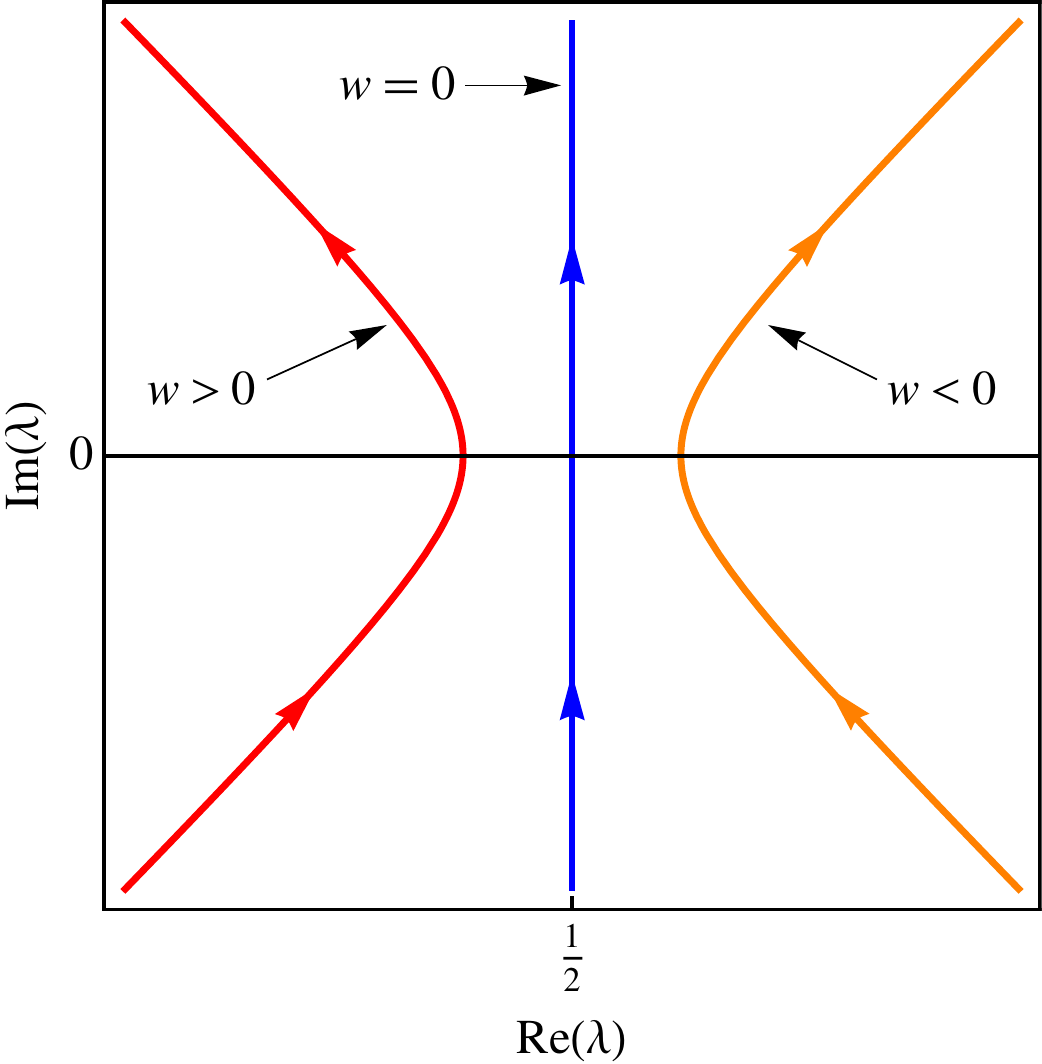} 
\caption{\label{sd-figure}(Color
  online). The paths of the steepest descent for $w>0$ (red), $w=0$
 (blue) and $w<0$ (orange) respectively.}
\end{figure}

We have found in the previous subsection that the saddle-point is
located at $\lambda^*\in(\lambda_-,\lambda_+)$ given
by \eref{lambda*}. We intend to deform the original contour of
integration in \eref{PDF} into the steepest descent path that passes
through $\lambda^*$ and the imaginary part of $f_w(\lambda)$ is
constant along the new path.  Setting $\mathrm{Im}
[f_w(\lambda)]=\mathrm{constant} =\mathrm{Im} [f_w(\lambda^*)] =0$
gives the path of the steepest descent as (see \fref{sd-figure})
\begin{equation}
\lambda_R=\frac{1}{2} \left[1-\frac{w}{\sqrt{\alpha}} \sqrt{\frac{1+\alpha}{w^2+\alpha}+4\lambda_I^2} \right]
\label{sd-path}
\end{equation}
 where $\lambda_R=\mathrm{Re}(\lambda)$ and
$\lambda_I=\mathrm{Im}(\lambda)$ are, respectively, the real and
imaginary parts of $\lambda$, that is, $\lambda=\lambda_R +
i \lambda_I$. The steepest descent path intersects the real-$\lambda$
axis at $\lambda^*$ at an angle $\phi=\pi/2$, which is evident
from \eref{f2} as well as \eref{sd-path}.

\subsubsection{Case: $\alpha \le 1/3$}

We have found in \sref{LDF} that $g(\lambda)$ does not have any
singularities in $(\lambda_-,\lambda_+)$ for $\alpha<1/3$. Therefore,
we can smoothly deform the contour along the path of the steepest
descent through the saddle-point. Subsequently, following the the
usual saddle-point approximation method, we write
\begin{equation}
 \int_{-i\infty}^{+i\infty} g(\lambda)\, e^{\tau f_w(\lambda)}\,
d\lambda \approx\frac{\sqrt{2\pi} g(\lambda^*) e^{\tau f_w(\lambda^*)}
e^{i\phi}}{|\tau f''_w(\lambda^*)|^{1/2}}.
\label{saddle-point approximation}
\end{equation}
Using Eqs.~\eqref{g(lambda)}, \eqref{f_w}, \eqref{eta-lambda-star},
\eqref{h1(w)-0}, \eqref{lambda0}, \eqref{f2}, and \eqref{h2(w)}, with
some algebra, we find that
\begin{align}
K(w)&\equiv\frac{\sqrt{2}\,g(\lambda^*)}{\sqrt{|f''_w(\lambda^*)|}} \notag\\
&= \alpha^{3/2} (1+1/\alpha)^{3/4} (w^2+\alpha)^{-5/4} \notag\\
&\times \Bigl[1+(w+\alpha) \lambda^*(w) - h_1(w)\Bigr]^{-1}\notag\\
&\times \Bigl[\bigl\{h_2(w)-h_1(w)\bigr\} +
(w-\alpha) \bigl\{\lambda^*(w) -\lambda_0\bigr\}\Bigr]^{-1}.
\label{K(w)}
\end{align}
It should be noted that, although the last line in the above
expression is written in that particular fashion involving $\lambda_0$
and $h_2(w)$, instead of $[1+(w-\alpha) \lambda^*(w) - h_1(w)]^{-1}$,
it does not have any singularities for $\alpha < 1/3$ as we have
discussed in the previous subsection. Even for $\alpha=1/3$, the above
expression diverges only in the limit of $w\rightarrow-\infty$.


\begin{figure*} 
\includegraphics[width=.32\hsize]{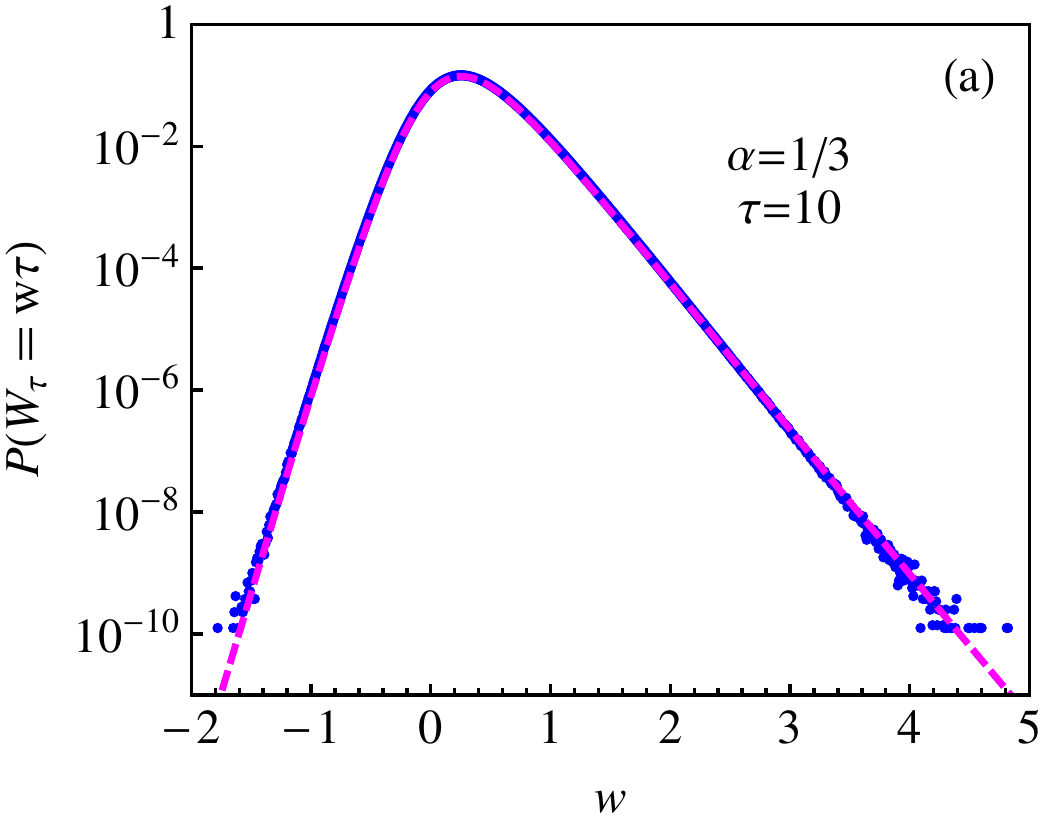} 
\includegraphics[width=.32\hsize]{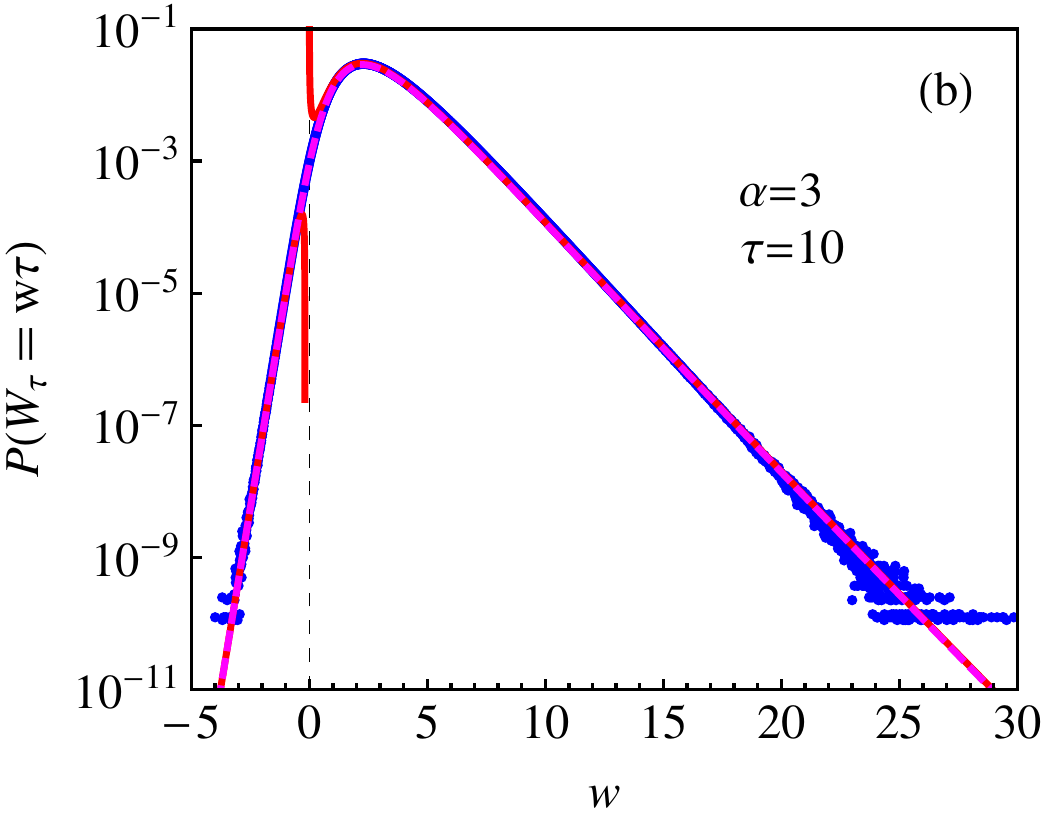} 
\includegraphics[width=.32\hsize]{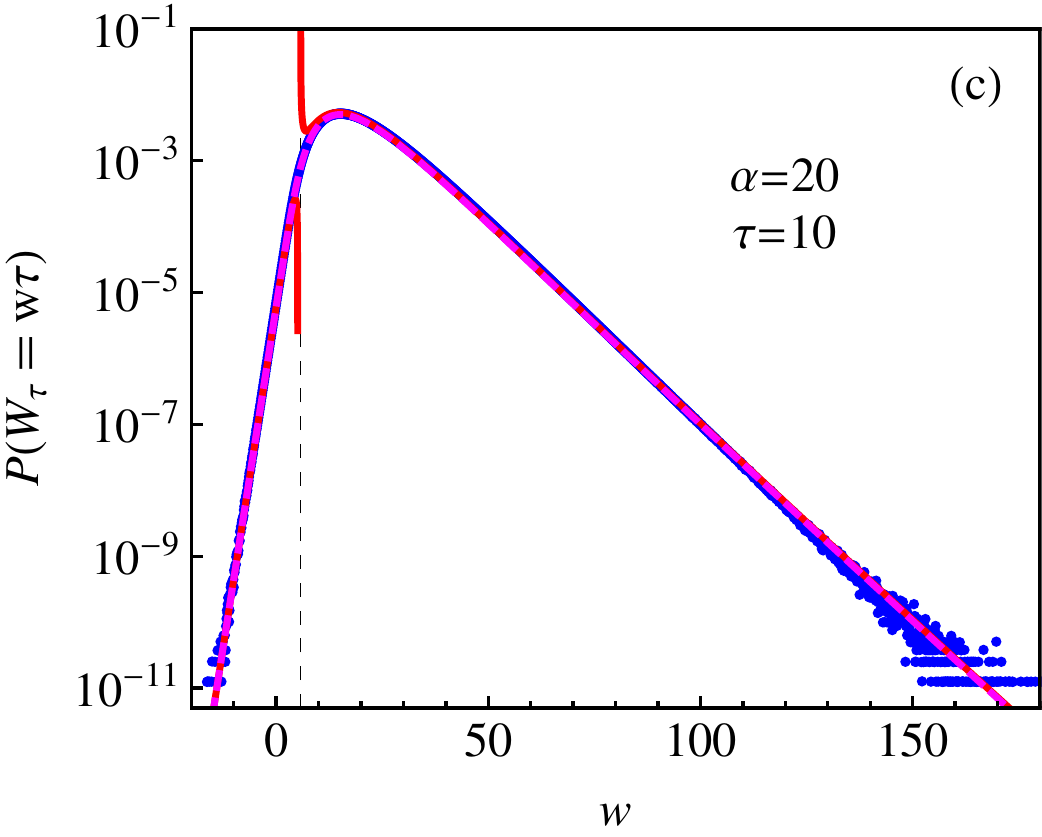} 

\caption{\label{PDF-figure}(Color online). $P(W_\tau)$ against the scaled variable $w=W_\tau/\tau$ for $\tau=10$ and (a) $\alpha=1/3$, (b) $\alpha=3$, and (c)
 $\alpha=20$, respectively. The points (blue) are obtained from
 numerical simulation, and the dashed solid lines (magenta) plot the
 analytical asymptotic forms given by \eref{PDF-1} for (a)
 and \eref{PDF-general} for (b) and (c).  In (b) and (c): the solid
 (red) lines plot the form given by \eref{PDF-2}, and the dashed
 vertical lines mark the positions of $w=w^*$.}
\end{figure*}


Finally, using \eref{saddle-point approximation}, and \eref{K(w)},
in \eqref{PDF}, we find that (recall that $\phi=\pi/2$)
\begin{equation}
P(W_\tau=w\tau) \approx \frac{K(w)}{2\sqrt{\pi \tau}}\, e^{\tau
h_1(w)}.
\label{PDF-1}
\end{equation}
It is seen from \fref{PDF-figure}(a) that the above form is a perfect
fit for the data obtained using numerical simulation for $\alpha=1/3$.

\subsubsection{Case: $\alpha \ge 1/3$}

Let us consider the case $\alpha \ge 1/3$. We have found in \sref{LDF}
that $g(\lambda)$ for this case has a pole at $\lambda_0 \in
(0,\lambda_+]$ and $\lambda^*(w) < \lambda_0$ for $w > w^*$, whereas
$\lambda^* (w) >\lambda_0$ for $w< w^*$.
Now if $w \gg w^*$, then the asymptotic behavior of \eref{PDF} is
obtained from the saddle-point approximation, which is given
by \eref{PDF-1}.
On the other hand, when $w \ll w^*$, while deforming the contour of
the integration in \eref{PDF} to the steepest descent path through the
saddle point, it goes around the pole at $\lambda_0$ in a clockwise
direction. Therefore, according to the residue theorem, the integral
pick up a contribution from the pole which is given by
\begin{math}
-g_{-1}\, \exp[{\tau f_w(\lambda_0)}], 
\end{math}
where
\begin{equation}
g_{-1}=\lim_{\lambda\rightarrow\lambda_0} \bigl[(\lambda-\lambda_0)\,
g(\lambda) \bigr]
=-\frac{(3\alpha-1)^2}{8\alpha^2(1+\alpha)}~.
\label{g-1}
\end{equation}
It is convenient to use \eref{g(lambda)-2} for $g(\lambda)$ to obtain
the last expression.  The saddle-point contribution to the integral is
same as given by \eref{PDF-1}.  Therefore, combining the contributions
from the saddle-point as well as from the pole, we can write that for
$\alpha \ge 1/3$ when $|w-w^*| \gg 0$,
\begin{align}
P(W_\tau=w\tau) &\approx  \frac{K(w)}{2\sqrt{\pi \tau}}\, e^{\tau
h_1(w)} \notag\\
&-\theta(w^*-w)\,g_{-1}\, e^{\tau
h_2(w)}.
\label{PDF-2}
\end{align}
Since $h_1(w^*)=h_2(w^*)$ and $\lambda^*(w^*)=\lambda_0$, it is clear
from \eref{K(w)} that $K(w)$ diverges as $w\rightarrow
w^*$. Therefore, \eref{PDF-2} does not provide the correct description
of the actual probability density function as $w$ approaches $w^*$ (from any side), which is
also seen from \fref{PDF-figure}(b) and \fref{PDF-figure}(c).

We carry out an asymptotic analysis of the integral along the path of
the steepest descent, which is also valid when $w$ is near $w^*$,
using the method of uniform asymptotic expansions~\cite{Wong}.  The
steps are outlined in Appendix~\ref{uniform asymptotic}, following
which we get
\begin{widetext}
\begin{equation}
P(W_\tau=w\tau) \approx  \frac{ e^{\tau
h_1(w)}}{2\sqrt{\pi \tau}}\left[K(w)
-\frac{\sgn{w^*-w}\,g_{-1}}{\sqrt{h_2(w)-h_1(w)}}\right]
+  e^{\tau h_2(w)}g_{-1} \left[
 \frac{\sgn{w^*-w}}{2}\, \erfc \bigl(\sqrt{\tau[h_2(w)-h_1(w)]}\,\bigr)
 -\theta(w^*-w)
\right].
\label{PDF-general}
\end{equation}
\end{widetext}
Again, this above asymptotic form matches with simulation results
extremely well even for $\tau=10$ as seen from \fref{PDF-figure}(b)
and \fref{PDF-figure}(c).
It follows from Appendix~\ref{uniform asymptotic} that the above
equation reduces to \eref{PDF-2} for $|w-w^*| \gg 0$.
\Eref{PDF-general}  is valid for any values of $w$ including $w=w^*$.

\section{Remarks on stochastic integration}
\label{stochastic integration}

\begin{figure}
\includegraphics[width=.9\hsize]{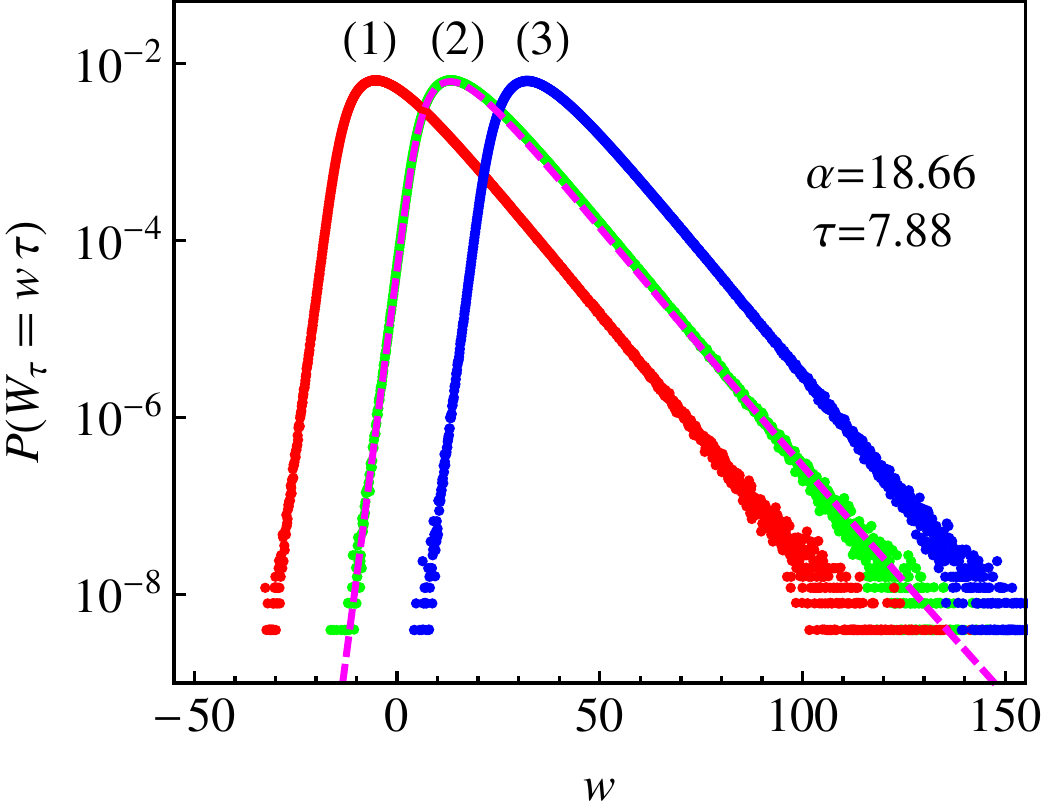}
\caption{\label{integration-scheme} (Color online) The three sets [(1) red, (2) green and
(3) blue] of points plot the probability density functions of the work
fluctuations obtained from numerical simulations of the Langavin
equation given by \eref{Langevin-HO}, while using the three different
numerical integration schemes respectively, outlined
in \sref{stochastic integration} for the work integral given
by \eref{Work}.  The dashed line (magenta) plots the analytical
asymptotic forms given by \eref{PDF-general}. It is evident that only
the second stochastic integration scheme of \sref{stochastic
integration} yields the correct distribution.}
\end{figure}

Finally, we make some remarks on the evaluation of the integrals of
type $\int_0^\tau f(t) v(t) dt$ --- that appears in Eqs.~\eqref{heat}
and \eqref{Work} --- in numerical simulations or from experimental
data. First the total time interval $(0,\tau)$ is divided into small
time intervals of $\Delta t$ duration such that
\begin{equation}
\int_0^\tau f(t) v(t) dt = \sum_{n=0}^{N-1} \int_t^{t+\Delta t} f(t')
v(t') dt',
\end{equation}
where $t=n\Delta t$ and $N\Delta t=\tau$. Next the integral on the
right side of the above equation needs to be evaluated for each
interval $(t,t+\Delta t)$. At first glance, it might look reasonable
to assume that the integrals could be evaluated numerically in
simulations or from experimental data by any one of the following
approximation schemes:

(1)~$\displaystyle\int_t^{t+\Delta t} f(t') v(t') dt' \approx v(t) F_{\Delta t} (t) $,

(2)~$\displaystyle\int_t^{t+\Delta t} f(t') v(t')
dt' \approx \frac{1}{2}[v(t) + v(t+\Delta t)] F_{\Delta t} (t) $,

(3)~$\displaystyle\int_t^{t+\Delta t} f(t') v(t') dt' \approx
v(t+\Delta t) F_{\Delta t} (t) $,
\\
where $F_{\Delta t} (t) = \int_t^{t+\Delta t} f(t') dt'$, which is of
$O(\sqrt{\Delta t})$.  However, it turns out that the second one (\`a
la Stratonovich) is the only correct scheme to follow.  If one were to
use the first scheme instead, the resulting distribution would shift
to the left of the true one, whereas the use of the third scheme would
result in a shift to the right of the actual distribution
(see \fref{integration-scheme}).  While comparing our asymptotic
form \eref{PDF-general} with the experimental distribution of
Ref.~\cite{Ciliberto:10}, the work is evaluated using the second
scheme given above~\cite{SS:11}.  The second scheme is also used while
deriving the Fokker-Planck equation given by
Eqs.~\eqref{FP-eq} and \eqref{FP-op}.  On the other hand, the methods
used here to obtain the solution of the Fokker-Planck equation does
not require any such stochastic integration scheme~\cite{Kundu:11}.
We have used the Fokker-Planck equation to merely verify the
solution. This also proves the correctness of the second scheme in the
present context.

\section{Summary}
\label{summary}

In this paper, we have applied the recent formalism of
Kundu \textit{et al.}~\cite{Kundu:11} to the case of a harmonically
bound Brownian particle coupled to two heat baths at different
temperatures. We have considered the fluctuations of the total amount
of heat flow $Q$ from one of the baths to the particle in a given time
duration $\tau$.  Its characteristic function for given initial and
final phase-space configurations satisfies a non-trivial Fokker-Planck
equation. We have obtained the largest eigenvalue of the Fokker-Planck
operator as well as the corresponding eigenfunctions (left and right)
exactly. Using those, and integrating over the final configurations
and averaging over the initial configurations with respect to the
nonequilibrium steady-state, we have obtained the characteristic
function $\langle e^{-\lambda Q}\rangle\approx g(\lambda)\exp[\tau\mu
(\lambda)]$ for large $\tau$, where the both $\mu(\lambda)$ and
$g(\lambda)$ have been obtained exactly.
A special case of this problem corresponds to the work done by an
external stochastic force on a harmonic oscillator coupled to a
thermal bath. 
This special case in fact models an recently studied experimental
system of a stochastically driven atomic-force microscopy
cantilever~\cite{Ciliberto:10}.
For this model we have analyzed the characteristic function and found
the exact large deviation function as well as the complete asymptotic
forms of the probability density function of the work.  We have
compared the analytical results with numerical simulations and found
excellent agreements between the theory and simulation. In fact, the
theoretical asymptotic forms of the probability density function
also match quite well with the experimentally obtained
forms~\cite{SS:11}.
Finally, we believe that the results as well as the analytical methods
of this paper would be useful for many other similar problems.

\begin{acknowledgements}
I thank Abhishek Dhar for useful discussions and acknowledge the
support of the Indo-French Centre for the Promotion of Advanced
Research (IFCPAR/CEFIPRA) under Project 4604-3.
\end{acknowledgements}

\appendix

\section{Fluctuations of the heat flow in a harmonic chain}
\label{Harmonic chain}

Here, for ease of reference, we outline the relevant results, of
Ref.~\cite{Kundu:11}, for a harmonic chain consisting of $N$ particles
connected at its two ends to white-noise Langevin reservoirs at
different temperatures $T_L$ and $T_R$ respectively.
The system, described by the variables ${X}^T=(x_1,x_2,\dots,x_N)$ and
${V}^T=(v_1,v_2,\dots,v_N)$, evolves according to the following
equations of motion:
\begin{equation}
\dot{X}={V},\quad
{\bm M} \dot{{V}}=-\bm{\Phi} X -\bm{\gamma} V + \eta (t),
\label{Langevin}
\end{equation}
where $\bm{M}=\text{diag} (m_1,m_2,\dots,m_N)$ is the mass matrix,
$\bm{\Phi}$ is the force matrix, $\bm{\gamma}$ is the dissipation
matrix with elements
${\bm \gamma}_{i,j}=\delta_{i,j}(\delta_{i,1}\gamma_L
+\delta_{i,N}\gamma_R)$, and ${\eta}$ is the Gaussian noise vector
with elements ${\eta}_i(t)= \delta_{i,1}\eta_L(t)
+\delta_{i,N}\eta_R(t)$ whose correlators are given by
Eqs.~\eqref{correlator-1}-\eqref{correlator-3}.
The quantity of interest is the total amount of heat flowing from one
of the reservoirs --- say the left (L) --- into the system in a given
time duration $\tau$, given by
\begin{equation}
  Q=\int_0^\tau \bigl[\eta_L(t) - \gamma_L v_1(t)\bigr] v_1(t)\; dt.
\end{equation}

It was found in Ref.~\cite{Kundu:11} that the restricted
characteristic function $Z(\lambda,U,\tau|U_0)=\bigl \langle
e^{-\lambda Q}\, \delta[U-U(\tau)] \bigr\rangle_{U_0}$ for fixed
initial and final configurations, $U_0^T=(X_0^T,V_0^T)$ and
$U^T=(X^T,V^T)$,  respectively, has the large-$\tau$ asymptotic form
\begin{equation}
  Z(\lambda,U,\tau |U_0)
\sim
  \chi({U}_0,\lambda)\Psi({U},\lambda)\,
  \exp[\tau\mu(\lambda)].
\end{equation}
The cumulant generating function $\mu(\lambda)$ is given by
\begin{equation}
  \mu(\lambda)=-\frac{1}{4\pi}\int_{-\infty}^\infty
  d\omega \ln \Bigl[1+
\lambda(\Delta\beta-\lambda) \mathcal{T}_1(\omega) \Bigr],
\label{mu(lambda)}
\end{equation}
where $\Delta\beta=(\gamma_R/d_R)-(\gamma_L/d_L)$ and
\begin{align}
\label{T(omega)}
\mathcal{T}_1(\omega)&=4 d_L d_R
\omega^2\bm{G}^+_{1,N}(\omega)\bm{G}^-_{1,N}(\omega),\\
\intertext{with}
\label{G}
\bm{G}^\pm(\omega)&=\Bigl[\bm{\Phi}
-\omega^2\bm{M}\pm i\omega\bm{\gamma}\Bigr]^{-1}~.
\end{align}
The above formula of $\mu(\lambda)$ has been also generalized to the
case of heat conduction across arbitrary harmonic
networks~\cite{Saito:11}.

The functions $\Psi(U,\lambda)$ and $\chi(U_0,\lambda)$ have the
following Gaussian forms:
\begin{align}
\label{wave function}
 \Psi(U,\lambda) &= \frac{\exp\left[-\frac{1}{2}
 U^T \bm{L}_1(\lambda) U\right]}{(2\pi)^{N} \sqrt{\det \bm{H}_1(\lambda)}}, \\
\label{projection}
 \chi(U_0,\lambda) &= \exp\left[-\frac{1}{2} U^T_0 \bm{L}_2(\lambda)
U_0\right], \\
\intertext{where}  
\label{L1}
\bm{L}_1(\lambda) &=\bm{H}_1^{-1}+ \bm{H}_1^{-1} \bm{H}_2^T, \\
\label{L2}
\bm{L}_2(\lambda) &=-\bm{H}_1^{-1} \bm{H}_2^T,
\end{align}
and $\bm{H}_1$ and $\bm{H}_2$ satisfy the relation
\begin{equation}
\bm{H}_2 \bm{H}_1^{-1} \bm{H}_2^T + \bm{H}_1^{-1}
   \bm{H}_2^T= \bm{H}_3.
\label{condition-H}
\end{equation}
The matrices $\bm{H}_1$, $\bm{H}_2$, and $\bm{H}_3$ are, respectively,
given as follows:
\begin{widetext}
\begin{align}
\label{H1}
 \bm{H}_1 (\lambda) &= \frac{1}{2} (~\bm{I}_1+\bm{I}_1^T~)
\quad\text{with}\quad\bm{I}_1 (\lambda)=
\frac{d_Ld_R}{\pi}\int_{-\infty}^{\infty} d\omega~ \frac{C_{1,1}F_2
  F_2^{\dagger} + C_{1,2}F_3 F_2^{\dagger} +C_{2,1}F_2 F_3^{\dagger} +
  C_{2,2}F_3 F_3^{\dagger}} {1+\lambda(\Delta \beta -\lambda)
  \mathcal{T}_1(\omega)},\\
\label{H2}
\bm{H}_2(\lambda) &=
\lim_{\epsilon\rightarrow 0}
\frac{\lambda}{\pi}
\int_{-\infty}^{\infty}
d\omega\,
 e^{i \omega \epsilon}\,
\frac{d_L (1-2i\omega \gamma_L
    {\bm{G}^+_{1,1}} )F_1^* F_2^{\dagger} 
-2i\omega (\gamma_L + \lambda d_L)d_R\,\bm{G}^+_{1,N}F_1^* F_3^{\dagger}
}{1+\lambda(\Delta \beta -\lambda) \mathcal{T}_1(\omega)},\\
\label{H3}
\bm{H}_3(\lambda) &=\frac{1}{2} (~\bm{I}_3+\bm{I}_3^T~)
\quad\text{with}\quad
\bm{I}_3(\lambda) = \frac{\lambda \bigl ( \gamma_L + \lambda d_L
  \bigr)}{\pi}\int_{-\infty}^{\infty} d\omega~\frac{F_1F_1^{\dagger}
}{1+\lambda(\Delta \beta -\lambda) \mathcal{T}_1(\omega)}.
\end{align}
In the above expressions $F_1$, $F_2$, and $F_3$ are column matrices,
given respectively by
\begin{align}
F_1^T&= \Bigl(
[\bm{G}^+\bm{\Phi}]_{1,1},[\bm{G}^+\bm{\Phi}]_{1,2},\dots,[\bm{G}^+\bm{\Phi}]_{1,
N},
-i\omega [\bm{G}^+\bm{M}]_{1,1},-i \omega
[\bm{G}^+\bm{M}]_{1,2},\dots,-i \omega [\bm{G}^+\bm{M}]_{1,N} 
\Bigr),
\label{F1def}\\
 F_2^T&= \Bigl(
{G}^+_{1,1},~{G}^+_{2,1},~\dots,{G}^+_{N,1},~
i\omega {G}^+_{1,1},~i \omega {G}^+_{2,1},\dots,i \omega
{G}^+_{N,1}~ \Bigr), 
\label{F2def}\\
 F_3^T&= \Bigl(
{G}^+_{1,N},~{G}^+_{2,N},\dots,{G}^+_{N,N},~
i\omega {G}^+_{1,N},~i \omega{G}^+_{2,N},\dots,i\omega
{G}^+_{N,N}\Bigr),
\label{F3def}
\end{align}
and $C_{i,j}$ is the $\{i,j\}$-th element of the matrix
\begin{equation}
\bm{C} (\omega) =
\left(\begin{array}{cc}\displaystyle
\frac{1}{d_R}  -4 \lambda\gamma_L \omega^2|{G}^+_{1,N}|^2 
&\displaystyle\quad 4 \lambda\gamma_L \omega^2
{G}^+_{1,1}{G}^-_{1,N}  + 2i \lambda\omega {G}^-_{1,N} \\[3mm]
\displaystyle
4\lambda \gamma_L \omega^2
{G}^-_{1,1}{G}^+_{1,N}  -2i\lambda \omega {G}^+_{1,N} 
&\displaystyle\frac{1}{d_L} + 4\lambda \gamma_R \omega^2 |{ G}^+_{1,N}|^2  
\end{array}\right).
\label{C_n}
\end{equation}

\end{widetext}

\subsection{The $N=1$ case}
\label{N=1 case}

Here we explicitly evaluate the above expressions for the special case
of the single Brownian particle in a harmonic potential that is
discussed in
\sref{Brownian particle}. For this case, we have
\begin{equation}
G^\pm(\omega)=\Bigl[k -\omega^2 m \pm i\omega \gamma\Bigr]^{-1},
\end{equation}
which gives
\begin{equation}
\mathcal{T}_1(\omega)=\frac{4 d_L d_R \omega^2}{(k-m\omega^2)^2
+ \gamma^2\omega^2}.
\end{equation}

In this case \eref{mu(lambda)} can be evaluated explicitly and the
final expression is given by \eref{mu(lambda) N=1}.

To obtain the explicit forms of of the functions given by
Eqs.~\eqref{wave function} and \eqref{projection}, we first find that
\begin{equation}
F_1=G^+ \binom{k}{-i \omega m}\quad\text{and}\quad
F_2=F_3=G^+ \binom{1}{i \omega}.
\end{equation}
Moreover, adding the elements of $\bm{C}(\omega)$ from \eref{C_n} and
 then using the identity $ -i [G^- - G^+] = 2\gamma \omega |G^+|^2$ we
 get $ C_{1,1} + C_{1,2} + C_{2,1} + C_{2,2}
 = d_L^{-1} + d_R^{-1}$.

Next, we carry out the integrations in Eqs.~\eqref{H1}-\eqref{H3} to find
that
\begin{align}
\bm{H}_1(\lambda)&=\frac{d_L + d_R}{\gamma\eta(\lambda)}
\left(\begin{array}{cc}
1/k &0\\
0 & 1/m
\end{array}
\right),\\[2mm]
\bm{H}_2(\lambda)&=\frac{\lambda d_L - (\gamma/2) [\eta(\lambda)-1]}{\gamma\eta(\lambda)}
\left(\begin{array}{cc}
1 &0\\
0 & 1
\end{array}
\right), \\[2mm]
\bm{H}_3(\lambda)&=\frac{\lambda(\gamma_L +\lambda d_L)}{\gamma\eta(\lambda)}
\left(\begin{array}{cc}
k &0\\
0 & m
\end{array}
\right).
\end{align}
It can be checked that the above matrices satisfy the relation given
by \eref{condition-H}.

Using the above matrices in Eqs.~\eqref{L1} and \eqref{L2},
respectively, we obtain
\begin{align}
\bm{L}_1 (\lambda)= B_+(\lambda)
\left(\begin{array}{cc}
k &0\\
0 & m
\end{array}
\right),\\
\bm{L}_2 (\lambda)= B_-(\lambda)
\left(\begin{array}{cc}
k &0\\
0 & m
\end{array}
\right),
\end{align}
where
\begin{equation}
B_\pm(\lambda)=
\frac{\gamma \eta(\lambda)\pm (\gamma + 2 \lambda d_L)}
{2 (d_L+d_R)}.
\label{B_pm}
\end{equation}

Finally, using the above results in
\eref{wave function} and \eref{projection}, we obtain \eref{psi}
and \eref{chi} respectively.


\section{Uniform asymptotic expansions for a saddle-point near a pole}
\label{uniform asymptotic}

Following Ref.~\cite{Wong}, we outline below the method of uniform
asymptotic expansion of a integral having a saddle-point near a pole.
Let us denote
\begin{equation}
I(\tau)  = \int_C g(\lambda)\,
e^{\tau f_w(\lambda)}\, d\lambda,
\label{I-1}
\end{equation}
where $C$ denotes the contour along the path of steepest descent.  We
set
\begin{equation}
f_w(\lambda) - f_w(\lambda^*) = -u^2.
\end{equation}
Since the imaginary part of $f_w(\lambda)$ is constant (which happens
to be zero in our particular case) on $C$, the above integral can be
converted into an integral with respect to a real variable $u$, as 
\begin{equation}
I(\tau) =  e^{\tau f_w(\lambda^*)}\int_{-\infty}^\infty q(u)\, e^{-\tau u^2}\, du,
\label{I-2}
\end{equation}
with
\begin{equation}
q(u)= g(\lambda) \,\frac{d\lambda}{du}.
\label{q(u)-1}
\end{equation}
The pole $\lambda_0$ of $g(\lambda)$ is then mapped to a pole of
 $q(u)$ at $u=ib$ with
\begin{equation}
b=\sgn{\lambda^*-\lambda_0}\sqrt{f_w(\lambda_0) - f_w(\lambda^*)}~.
\end{equation}
Note that, $f_w(\lambda)$ is minimum at $\lambda^*$ along
real-$\lambda$. Therefore, $f_w(\lambda_0) > f_w(\lambda^*)$, and
hence $b$ is real. We can write
\begin{equation}
q(u)=\frac{c_{-1}}{u - ib\, } + \psi (u),
\label{q(u)-2}
\end{equation}
where $c_{-1}=\lim_{u\rightarrow ib} \bigl[ (u-ib)\, q(u)\bigr]$, and $\psi(u)$ has
no pole at $u=ib$ or $u=0$. To evaluate $c_{-1}$ we note that
$u^2+b^2=f_w(\lambda_0) -f_w(\lambda)$ and $d\lambda/du=-2
u/f'(\lambda)$. Moreover, $f_w(\lambda)= f_w(\lambda_0)
+(\lambda-\lambda_0) f'(\lambda_0) +\dotsb$ as $\lambda\rightarrow\lambda_0$.
Therefore, we get
\begin{align}
c_{-1} &= \lim_{u\rightarrow
ib} \left[\left(\frac{u^2+b^2}{u+ib}\right) \left(\frac{d\lambda}{du}\right)
g(\lambda)\right] \notag\\ &=\lim_{\lambda\rightarrow\lambda_0}
\bigl[(\lambda-\lambda_0)\, g(\lambda)\bigr] =g_{-1},
\end{align}
where $g_{-1}$ is given by \eref{g-1}.

Let us first look at the integral
\begin{equation}
J_1(\tau) = \int_{-\infty}^\infty \frac{e^{-\tau u^2}}{u-ib}\, du.
\end{equation}
It satisfies the differential equation
\begin{equation}
J'_1(\tau) -b^2 J_1(\tau) = -ib \sqrt{(\pi/\tau)},
\end{equation}
with the boundary condition $J_1(\infty)=0$. It can be verified that the
solution is given by
\begin{equation}
J_1(\tau)= \sgn{b}\,i\pi\, e^{\tau b^2} \erfc(\sqrt\tau |b|).
\label{J1}
\end{equation}

Next, we look at the integral
\begin{equation}
J_2(\tau) = \int_{-\infty}^\infty \psi(u)\, e^{-\tau u^2}\, du.
\end{equation}
Expanding $\psi(u)$ Taylor series about $u=0$ $\psi(u)
=\sum_{n=0}^\infty \psi^{(n)} (0) u^n/n!$ in the above integral, and
then term by term integration of the series yields
\begin{equation}
J_2(\tau)= \sum_{m=0}^\infty \frac{\psi^{(2m)}(0)}{(2m)!} \frac{\Gamma(m+1/2)}{\tau^{m+1/2}}.
\label{J2-1}
\end{equation}
In terms of $J_{1,2}(\tau)$ we finally have
\begin{equation}
I(\tau)= e^{\tau f_w(\lambda^*)} \bigl[g_{-1} J_1(\tau)
+J_2(\tau)\bigr].
\label{I-3}
\end{equation}

In the following, we consider the large $\tau$ limit and find the
leading order contribution of $J_2(\tau)$, i.e.,
\begin{equation}
J_2(\tau)=\psi(0) \sqrt{(\pi/\tau)} + O\bigl(\tau^{-3/2}\bigr).
\label{J2-2}
\end{equation}
From \eref{q(u)-2} we get
\begin{equation}
\psi(0) = q(0) + \frac{g_{-1}}{ib},
\label{psi(0)}
\end{equation}
and then from \eref{q(u)-1} 
\begin{equation}
q(0)=
g(\lambda^*) \,\frac{d\lambda}{du}\bigg|_{\lambda\rightarrow\lambda^*}.
\label{q(0)-1}
\end{equation}
Near $\lambda^*$ we have 
\begin{equation}
-u^2=\frac{1}{2} f''(\lambda^*)
(\lambda-\lambda^*)^2 +\dotsb.
\end{equation}
Let $\lambda-\lambda^*= r e^{i\phi}$ and
$f''(\lambda^*)=|f''(\lambda^*)|e^{i\theta}$.  Therefore,
\begin{equation}
-u^2=\frac{r^2}{2} |f''(\lambda^*)|
e^{i(\theta+2\phi)} +\dotsb.
\end{equation}
Since, $u$ is real, we have $\theta+2\phi=\pi$, and
\begin{equation}
u=\pm\frac{r}{\sqrt2} |f''(\lambda^*)|^{1/2}
 +\dotsb.
\end{equation}
Thus,
\begin{equation}
\frac{du}{d\lambda}\bigg|_{\lambda\rightarrow\lambda^*}
= \frac{du}{dr} \frac{dr}{d\lambda}\bigg|_{\lambda\rightarrow\lambda^*}
=\frac{1}{\sqrt2} |f''(\lambda^*)|^{1/2} \, e^{-i\phi}.
\end{equation}
Substitution of this into \eref{q(0)-1} yields
\begin{equation}
 q(0)= \frac{\sqrt{2}\,g(\lambda^*)}{\sqrt{|f''_w(\lambda^*)|}}\, e^{i\phi}.
\label{q(0)-2}
\end{equation}
In our case, from \eref{f2} we get $\theta=0$, and therefore
$\phi=\pi/2$.

If the saddle-point and the pole are far apart, then using
$J_1(\tau) \sim (i/b) \sqrt{(\pi/\tau)}$ for large $b$ and $\tau$, it
is immediately checked that $I(\tau)$ reduces to the usual
saddle-point approximation given by \eref{saddle-point
approximation}. More generally, $I(\tau)$ given by \eref{I-3}, with
$J_{1,2}(\tau)$ given by Eqs.~\eqref{J1} and \eqref{J2-1}
respectively, is valid for all $b$, including the limit $b\rightarrow
0$. For large $\tau$, using Eqs.~\eqref{J2-2}, \eqref{psi(0)}
and \eqref{q(0)-2} we get 
\begin{equation}
J_2(\tau)=i \sqrt{\frac{\pi}{\tau}}
\left[
\frac{\sqrt{2}\,g(\lambda^*)}{\sqrt{|f''_w(\lambda^*)|}}
-\frac{g_{-1}}{b}
\right]
+ O\bigl(\tau^{-3/2}\bigr).
\end{equation}

\end{document}